\documentclass[aps,prc,twocolumn,showpacs,floatfix,superscriptaddress,amsmath,amssymb,nofootinbib,longbibliography]{revtex4-1}
\newcommand{\be}{\begin{equation}} \newcommand{\ee}{ \end{equation}}
\newcommand{\ba}{\begin{eqnarray}} \newcommand{\ea}{ \end{eqnarray}}
\newcommand{\ve}{\varepsilon} \usepackage{color}

\begin{document}

\title{Random-Matrix Approach to Transition-State Theory}

\author{H. A. \surname{Weidenm{\"u}ller}}
\email{haw@mpi-hd.mpg.de}

\affiliation{Max-Planck-Institut f{\"u}r Kernphysik, D-69029
  Heidelberg, Germany}

\begin{abstract}To model a complex system intrinsically separated by a
  barrier, we use two random Hamiltonians, coupled to each other
  either by a tunneling matrix element or by an intermediate
  transition state. We study that model in the universal limit of
  large matrix dimension. We calculate the average probability
  $\langle P_{a b} \rangle$ for transition from scattering channel $a$
  coupled to the first Hamiltonian to scattering channel $b$ coupled
  to the second Hamiltonian. Using only the assumption $\sum_{b'}
  T_{b'} \gg 1$ we find $\langle P_{a b} \rangle = P_a T_b / \sum_{b'}
  T_{b'}$. Here $P_a$ is the probability of formation of the tunneling
  channel or the transition state, and the $T_{b'}$ are the
  transmission coefficients for channels $b'$ coupled to the second
  Hamiltonian. That result confirms transition-state theory in its
  general form. For tunneling through a very thick barrier the
  condition $\sum_{b'} T_{b'} \gg 1$ is relaxed and independence of
  formation and decay of the tunneling process hold more generally.
  \end{abstract}

\maketitle

\section{Introduction}

Barrier penetration and transition over a barrier play important roles
in several areas of quantum physics. Since the pioneering work of Bohr
and Wheeler~\cite{Boh39}, transition-state theory has been an
important element in the theory of nuclear fission~\cite{Kra12}. It
also plays a key role in physical chemistry~\cite{Tru96}. In two
recent papers, Bertsch and Hagino~\cite{Ber21} and Hagino and
Bertsch~\cite{Hag21} have studied a statistical model for a
transition-state process wherein the complete mixing of states is
hindered by an internal barrier. The model consists of two
uncorrelated Hamiltonians, each a member of the Gaussian Orthogonal
Ensemble (GOE) of random matrices. The first random Hamiltonian is fed
from some entrance channel. Penetration through the barrier separating
the two Hamiltonians is due to a single channel. We believe that for
transition-state theory, such a random-matrix approach is of
considerable interest. Random-matrix theory is a universal tool for
modeling complex quantum systems~\cite{Guh98}. It exhibits generic
features of such systems without resorting to specific dynamical
assumptions. These features emerge in the limit of infinite matrix
dimension which allows for a clean separation of local fluctuation
properties (that are generic) and global properties (that are not).

The present paper is motivated by the fact that the model used in
Refs.~\cite{Ber21, Hag21} does not possess a realistic GOE limit of
large matrix dimension $N$. In brief, in the propagators of the model
Hamiltonians, the authors replace the energy $E$ by $E + i
\Gamma$. The constant offset $\Gamma$ is supposed to account for the
coupling to open channels.  Inspection shows that the number of
channels so simulated is equal to the matrix dimension $N$. If
$\Gamma$ is independent of $N$, the number of open channels tends to
infinity with $N$. It is easily checked that the fluctuations worked
out in Refs.~\cite{Ber21, Hag21} then tend to zero for $N \to
\infty$. That is unrealistic. The issue is presently addressed by the
authors~\cite{Ber22}. We do not go into detail because we aim at
presenting a viable alternative to the approach by Bertsch and
Hagino. We investigate a model which does possess a realistic GOE
limit with non-vanishing fluctuations.

Actually we study two simplified versions of the model of
Refs.~\cite{Ber21, Hag21}. Each employs two coupled GOE
Hamiltonians. In the first, the two GOE Hamiltonians are coupled by a
rank-one interaction. That is a model for barrier penetration via a
single channel. In the second, the two GOE Hamiltonians are coupled
via a single state. That models a physical transition state located
right above the barrier. In each case, we calculate the probability
for the transition from the entrance channel to some final channel on
the other side of the barrier. We formulate the conditions under which
that probability is the product of two or three statistically
uncorrelated factors. Then the decay following barrier penetration or
passage through the transition state is independent of its mode of
formation. Such independence is the hallmark of transition-state
theory. If there are sufficiently many open channels coupled
sufficiently strongly to the second Hamiltonian, we retrieve the
standard expression of transition-state theory.

The two versions of our model Hamiltonian are defined in
Section~\ref{mod}. The coupling to the channels and the scattering
matrix are defined in Section~\ref{scat}. In Section~\ref{asy} we
define the conditions of validity of transition-state theory within
our model and display the resulting transition probability. The
special case of a thick barrier is treated in
Section~\ref{thi}. Section~\ref{sum} contains a brief
summary. Technical details are deferred to an appendix.

\section{Two Hamiltonians}
\label{mod}

The tunneling Hamiltonian is
\ba
\label{t1}
H_{\rm tun} = \left( {\begin{array}{cc}
  H_1 & V \\
  V^T & H_2 \\ \end{array} } \right) \ .
\ea
The index $T$ denotes the transpose, and $V$ is a rank-one interaction
that models tunneling. The Hamiltonian which carries the transition
state with energy $E_0$ is
\ba
\label{t2}
H_{\rm tra} = \left( { \begin{array}{ccc}
    H_1 & V_1 & 0 \\
  V^T_1 & E_0 & V^T_2 \\
  0 & V_2 & H_2 \\ \end{array} } \right) \ .
\ea
Both $H_1$ and $H_2$ are coupled to the transition state by vectors
$V_1$ and $V_2$, respectively. In Eqs.~(\ref{t1}, \ref{t2}), $H_1$ and
$H_2$ are GOE matrices of dimension $N \gg 1$ each, defined in two
Hilbert spaces denoted as space~$1$ and space~$2$, respectively. The
elements of the matrices $H_1$ and $H_2$ are uncorrelated
zero-centered Gaussian random variables with second moments
\ba
\label{t3}
\langle (H_1)_{\mu_1 \mu'_1} (H_1)_{\mu_2 \mu'_2} \rangle &=&
\frac{\lambda^2}{N} (\delta_{\mu_1 \mu_2} \delta_{\mu'_1 \mu'_2} +
\delta_{\mu_1 \mu'_2} \delta_{\mu_2 \mu'_1}) \ ,
\nonumber \\
\langle (H_2)_{\nu_1 \nu'_1} (H_2)_{\nu_2 \nu'_2} \rangle &=&
\frac{\lambda^2}{N} (\delta_{\nu_1 \nu_2} \delta_{\nu'_2 \nu'_2} +
\delta_{\nu_1 \nu'_2} \delta_{\nu_2 \nu'_1}) \ .
\ea
Angular brackets denote the ensemble average. Greek letters $\mu,
\mu', \mu_1, \mu'_1$, $\rho, \rho'$ etc. range from $1$ to $N$ in
space~$1$ while $\nu, \nu', \nu_1, \nu'_1, \sigma, \sigma'$ etc. range
from $N + 1$ to $2 N$ in space~$2$. The center row and column of the
matrix~(\ref{t2}) carry the index zero. The spectra of $H_1$ and $H_2$
range from $- 2 \lambda$ to $+ 2 \lambda$. For both Hamiltonians, the
average level density as function of energy $E$ is given by
\ba
\label{t4}
\rho(E) = \frac{N}{\pi \lambda} \bigg(1 - [E / (2 \lambda)]^2
\bigg)^{1 / 2} \ .
\ea
The rank-one interaction $V$ in Eq.~(\ref{t1}) is given by
\ba
\label{t5}
V_{\mu \nu} = (O_1)_{\mu N} {\cal V} (O_2)_{(N + 1) \nu} \ .
\ea
Here ${\cal V}$ is a constant with dimension energy, and $(O_1)_{\mu
  N}$ ($(O_2)_{(N + 1) \nu}$) are the elements of the $N^{\rm th}$
column (of the first row, respectively) of two arbitrary
$N$-dimensional orthogonal matrices in space~$1$ (space~$2$,
respectively). Transforming
\ba
\label{t6}
H_{\rm tun} \to \left( { \begin{array}{cc}
  O^T_1 & 0 \\
  0 & O_2 \\ \end{array} } \right) H_{\rm tun}
\left( { \begin{array}{cc}
    O_1 & 0 \\
  0 & O^T_2 \\ \end{array} } \right)
\ea
and using the orthogonal invariance of the GOE leaves the ensembles
$H_1$ and $H_2$ unchanged but changes $V$ into
\ba
\label{t7}
V_{\mu \nu} = \delta_{\mu N} \delta_{\nu (N + 1)} {\cal V} \ .
\ea
For the Hamiltonian in Eq.~(\ref{t2}), the vectors $V_1 = \{ V_{1 \mu}
\}$ and $V_2 = \{ V_{2 \nu} \}$ are written, respectively, as
$\sqrt{\sum_{\mu'} V^2_{1 \mu'}} \ \{ e_{1 \mu} \}$ and as
$\sqrt{\sum_{\nu'} V^2_{2 \nu'}} \ \{ e_{2 \nu} \}$. Here $\{ e_{1 \mu}
  \}$ and $\{ e_{2 \nu} \}$ are unit vectors in $N$ dimensions. Each
  of these forms one column of an orthogonal $N$-dimensional matrix
  $O_1$ and $O_2$, respectively. Via a transformation similar to that
  in Eq.~(\ref{t6}), the vectors $V_1$ and $V_2$ take the form
\ba
\label{t8}
V_{1 \mu} = \delta_{\mu N} {\cal V}_1 \ , \ V_{2 \nu} =
\delta_{\nu (N + 1)} {\cal V}_2 \ .
\ea
Here
\ba
\label{t9}
\frac{1}{N} {\cal V}^2_1 &=& \frac{1}{N} \sum_\mu V^2_{1 \mu} =
\overline{V^2_1} \ , \nonumber \\ 
\frac{1}{N} {\cal V}^2_2 &=& \frac{1}{N} \sum_\nu
V^2_{2 \nu} = \overline{V^2_2} \ .
\ea
Eqs.~(\ref{t1}) and (\ref{t7}) on the one hand and Eqs.~(\ref{t2}),
(\ref{t8}) and (\ref{t9}) on the other define our two 
Hamiltonians. Without loss of generality we assume that ${\cal V}_1,
{\cal V}_2, {\cal V}$ are positive.

In Section~\ref{asy} we show that to be physically meaningful, ${\cal
  V}$ must be independent of $N$ and small compared to $\lambda$. An
estimate for ${\cal V}_1$ and ${\cal V}_2$ is obtained by introducing
the mean GOE level spacing $d = \pi \lambda / N$ at the center $E = 0$
of the GOE spectrum. Then ${\cal V}^2_1 / \lambda = (\pi / d)
\overline{V^2_1}$ and ${\cal V}^2_2 / \lambda = (\pi / d)
\overline{V^2_2}$. We use the standard expressions for the spreading
widths that account for the coupling of the transition state to the
states in space~$1$ (space~$2$) with mean square coupling elements
$\overline{V^2_1}$ ($\overline{V^2_2}$, respectively),
\ba
\label{t10}
\Gamma^\downarrow_1 = (2 \pi / d) \overline{V^2_1} \ , \
\Gamma^\downarrow_2 = (2 \pi / d) \overline{V^2_2} \ .
\ea
To be physically meaningful, $\Gamma^\downarrow_1$ and
$\Gamma^\downarrow_1$ must be substantially smaller than the range of
the GOE spectrum, and that same statement then applies to ${\cal V}_1$
and ${\cal V}_2$. We define dimensionless positive parameters
\ba
\label{t11}
\tilde{\cal V} = \frac{\cal V}{\lambda} \ , \ \tilde{\cal V}_1 =
\frac{{\cal V}_1}{\lambda} \ , \ \tilde{\cal V}_2 =
\frac{{\cal V}_2}{\lambda}
\ea
all of which are independent of $N$ and small compared to unity. In
what follows we think of the parameter $\tilde{\cal V}$ as being
determined by a semiclassical calculation of the transition
probability through the barrier. The parameters $\tilde{\cal V}_1$ and
$\tilde{\cal V}_2$ measure the strength of the coupling of the
transition state to spaces~$1$ and $2$ and can only be determined by a
fit to data.

When $H_1$ and $H_2$ each range over the orthogonal ensemble, the
tunneling matrix element ${\cal V}$ becomes distributed uniformly over
the states in space~$1$ and space~$2$. That follows from
Eq.~(\ref{t5}) as then the matrices $O_1$ and $O_2$ independently
range over the set of orthogonal matrices in $N$ dimensions.
Therefore, tunneling is equally likely from any state $\mu$ in
space~$1$ to any state $\nu$ in space~$2$.  Pictorially speaking, in
the model Hamiltonian~(\ref{t1}) the tunneling barrier is replaced by
a wall of uniform thickness that extends over the entire GOE
spectrum. In an actual physical situation, the tunneling barrier has
more or less the shape of an inverted parabola. Tunneling becomes ever
more likely as the energy of the system approaches the top of the
barrier from below. For the Hamiltonian~(\ref{t1}) we simulate that
feature by keeping the system's energy fixed at the center $E = 0$ of
the GOE spectrum and increasing ${\cal V}$. The top of the barrier is
approached for very large ${\cal V}$. The Hamiltonian~(\ref{t2})
continues that physical picture to energies above the barrier where
the transition process is enhanced by a resonance at energy $E_0$.

\section{Scattering Matrix and Transition \\ Probability}
\label{scat}

In the construction of the scattering matrix we follow
Ref.~\cite{Wei84}. The open channels labeled $a, a', \ldots$ ($b, b',
\ldots$) are coupled, respectively, to states in space~$1$ (in
space~$2$) by real coupling matrix elements $W_{a \mu}$ ($W_{b
  \nu}$). Because of the normalization of the scattering wave
functions, these matrix elements have dimension (energy)$^{1/2}$. The
number of channels coupled to either space~$1$ or space~$2$ is
finite. It is held fixed as we take the limit $N \to \infty$ of
infinite matrix dimension in the following Sections. In spaces~$1$ and
$2$ we define the finite-rank width matrices $\Gamma_1$ and $\Gamma_2$
with elements
\ba
\label{s1}
(\Gamma_1)_{\mu \mu'} &=& 2 \pi \sum_{a'} W_{a' \mu} W_{a' \mu'} \ , \nonumber \\
(\Gamma_2)_{\nu \nu'} &=& 2 \pi \sum_{b'} W_{b' \nu} W_{b' \nu'} \ .
\ea
For the Hamiltonians $H_{\rm tun}$ and $H_{\rm tra}$ we define,
respectively,
\ba
\label{s2}
\Gamma_{\rm tun} = \left( { \begin{array}{cc}
  \Gamma_1 & 0 \\
  0 & \Gamma_2 \\ \end{array} } \right) \ , \
\Gamma_{\rm tra} = \left( { \begin{array}{ccc}
  \Gamma_1 & 0 & 0 \\
  0 & 0 & 0 \\
  0 & 0 &\Gamma_2 \\ \end{array} } \right) \ .
\ea
The elements for scattering from channel $a$ to channel $b$ of the
scattering matrix $S$ are
\ba
\label{s3}
S_{a b} = - 2 i \pi \sum_{\mu \nu} W_{a \mu} (D^{- 1})_{\mu \nu} W_{\nu b} \ ,
\ea
with $D$ given by
\ba
\label{s4}
D_{\rm tun} &=& E - H_{\rm tun} + (i/2) \Gamma_{\rm tun} \ , \nonumber \\
D_{\rm tra} &=& E - H_{\rm tra} + (i/2) \Gamma_{\rm tra} \ .
\ea
Eqs.~(\ref{s3}, \ref{s4}) are completely general. In applications,
however, the incident channel $a$ consists of two fragments, each in
its ground state. The final channels $b$ consist of all pairs of
fragments either in their ground or in excited states. In the converse
reaction $b \to a$, channel $b$ consists of two fragments, each in its
ground state. The final channels $a$ consist of all pairs of fragments
either in their ground or in excited states.

We expand $D^{- 1}_{\rm tun}$ in powers of ${\cal V}$. The resulting
series is odd in ${\cal V}$. We separate the first or the last factor
${\cal V}$. The remaining series containing even powers of ${\cal V}$
can be resummed. We expand $D^{- 1}_{\rm tra}$ in powers of the
elements in rows and columns labeled zero, respectively and proceed
analogously. The resulting element $S_{a b}$ of the scattering matrix
is the product of three dimensionless factors,
\ba
\label{s5}
&& S_{a b} = - i \bigg( \sum_\mu W_{a \mu} [(E - H_1 + (i/2)
  \Gamma_1)^{- 1}]_{\mu N} \sqrt{2 \pi \lambda} \bigg) \nonumber \\
&& \times {\cal A} \bigg( \sqrt{2 \pi \lambda} \sum_\nu
      [(E - H_2 + (i/2) \Gamma_2)^{- 1}]_{(N +1) \nu} W_{\nu b} \bigg) \ ,
      \nonumber \\
\ea
with
\ba
\label{s6}
{\cal A}_{\rm tun} &=& \tilde{\cal V} ( 1 - \xi_2 \tilde{\cal V} \xi_1
\tilde{\cal V} )^{- 1} \ , \nonumber \\
{\cal A}_{\rm tra} &=& \tilde{\cal V}_1 \frac{\lambda}{E - E_0 -
{\cal V}_1 \xi_1 \tilde{\cal V}_1 - {\cal V}_2 \xi_2 \tilde{\cal V}_2}
\tilde{\cal V}_2 \ .
\ea
The dimensionless random variables $\xi_1, \xi_2$ are defined as
\ba
\label{s7}
\xi_1 &=& \lambda [(E - H_1 + (i/2) \Gamma_1)^{- 1}]_{N N} \ ,
\nonumber \\
\xi_2 &=& \lambda [(E - H_2 + (i/2) \Gamma_2)^{- 1}]_{(N + 1) (N + 1)} \ .
\ea
For the transition probability $P_{a b}$ from channel $a$ to channel
$b$, Eq.~(\ref{s5}) gives
\ba
\label{s8}
&& P_{a b} = \bigg| \sum_\mu W_{a \mu} [(E - H_1 + (i/2)
  \Gamma_1)^{- 1}]_{\mu N} \sqrt{2 \pi \lambda} \bigg|^2 | {\cal A} |^2
\nonumber \\
&& \ \times \bigg| \sum_\nu \sqrt{2 \pi \lambda} [(E - H_2
  + (i/2) \Gamma_2)^{- 1}]_{(N +1) \nu} W_{\nu b} \bigg|^2 \ .
\ea
For both Hamiltonians~(\ref{t1}) and (\ref{t2}), the first (last) of
the three factors gives the probability to enter (leave) the tunneling
process or the transition state. The factor ${\cal A}_{\rm tun}$ in
the first of Eqs.~(\ref{s6}) accounts for tunneling, with the
denominator accounting for repeated tunneling events, and
correspondingly for the factor ${\cal A}_{\rm tra}$ in the second of
Eqs.~(\ref{s6}) which accounts for passage through the transition
state.

Each of the three factors in Eq.~(\ref{s8}) is a random variable
depending, in that order, on $H_1$, on both $H_1$ and $H_2$, and on
$H_2$. Therefore, the three factors are correlated. The probability of
decay, given by the last factor, is not independent of the mode of
formation of the tunneling process (or of the transition state).
Averaging $P_{a b}$ over the two ensembles does not help because the
average does not factorize. That shows that the standard assumption of
transition-state theory does not hold in general. Transition-state
theory holds if and only if the fluctuations of either $\xi_1$ or
$\xi_2$ or both in the second factor in Eq.~(\ref{s8}) are negligible
in which case at least one of the three factors in Eq.~(\ref{s8})
becomes uncorrelated with the rest. We now investigate the conditions
for that to be the case.

\section{Asymptotic Results}
\label{asy}

We determine the fluctuations of $\xi_1$ and $\xi_2$ by calculating
mean values and variances\footnote{We use the term variance for the
  expression $\langle |\xi - \langle \xi \rangle|^2 \rangle$.} of
these quantities in the limit of infite matrix dimension
$N$. Eqs.~(\ref{s7}) show that $\xi_1 \leftrightarrow \xi_2$ by
simultaneously interchanging $N \leftrightarrow N + 1$, $H_1
\leftrightarrow H_2$ and $\Gamma_1 \leftrightarrow \Gamma_2$.
Therefore, we confine ourselves to $\xi_1$. We use the close formal
similarity of $\xi_1$ to the $S$-matrix describing scattering by the
GOE Hamiltonian $H_1$ coupled to channels $a, a''$. That matrix is
defined by~\cite{Wei84}
\ba
\label{a1}
&& S_{a a''} = \delta_{a a''} \\
&& \ \ - 2 i \pi \sum_{\mu \mu'} W_{a \mu} [(E - H_1 + (i/2)
  \Gamma_1)^{- 1}]_{\mu \mu'} W_{\mu' a''} \ . \nonumber
\ea
The central piece of $S_{a a''}$ is the propagator $(E - H_1 + (i/2)
\Gamma_1)^{- 1}$. That same propagator defines $\xi_1$ in the first of
Eqs.~(\ref{s7}). In Ref.~\cite{Ver85} mean value and variance of $S_{a
  a''}$ were calculated analytically for $N \to \infty$ with the help
of the supersymmetry approach. Adopting that calculation to the
present case we obtain explicit expressions for average and variance
of $\xi_1$. Details are given in the Appendix. For the sake of
completeness we mention that combining results of Refs.~\cite{Sav05}
and \cite{Fyo05} yields an analytic expression for the full
distribution of $\xi_1$.

The average, given by
\ba
\label{a2}
\langle \xi_1 \rangle = - i \ ,
\ea
has magnitude unity and is independent of $\Gamma_1$. The variance of
$\xi_1$ does depend upon $\Gamma_1$ and is given by the threefold
integral in Eq.~(\ref{x2}) that involves the transmission coefficients
\ba
\label{a3}
T_a = 1 - | \langle S_{a a} \rangle |^2 \ . 
\ea
These are defined in terms of the average diagonal elements of the GOE
scattering matrix~(\ref{a1}) and measure the strength of the coupling
of channel $a$ to space~$1$. For a single open channel with
transmission coefficient $T_a$ the variance diverges like $1 / T_a$
for $T_a \to 0$. That is incontrast to the variance of $S_{a a''}$ in
Eq.~(\ref{a1}) which is bounded from above by unitarity. The variance
of $\xi_1$ decreases as the number of channels and the strength of
their couplings to space~$1$ increase. If the sum of the transmission
coefficients obeys $\sum_{a'} T_{a'} \gg 1$ we may use the asymptotic
expansion of the threefold integral in inverse powers of $\sum_{a'}
T_{a'}$ given in Ref.~\cite{Ver86}. As shown in the Appendix, the term
of leading order is
\ba
\label{a4}
\langle |\xi_1|^2 \rangle - 1 = \frac{2}{\sum_{a'} T_{a'}} \ {\rm for}
\ \sum_{a'} T_{a'} \gg 1 \ .
\ea
If the inequality in Eq.~(\ref{a4}) applies, $\xi_1$ in
Eqs.~(\ref{s6}) may be replaced by $\langle \xi_1 \rangle$, and
transition-state theory applies. In practice, the fluctuations of
$\xi_1$ are sufficiently small if the number of open channels with
transmission coefficients of order unity coupled to $H_1$ is of order
$10$ or bigger.

If the inequality in Eq.~(\ref{a4}) holds, the first factor on the
right-hand side of Eq.~(\ref{s8}) is not correlated with the rest. It
is then meaningful to calculate mean value and variance of the
amplitude (the expression under the absolute sign). Using the same
steps as for $\xi_1$ we show in the Appendix that the mean value of
the amplitude vanishes and that for $\sum_{a'} T_{a'} \gg 1$ the
variance, given by the leading term in the asymptotic expansion in
inverse powers of $\sum_{a'} T_{a'}$, is equal to
\ba
\label{a5}
&& 2 \pi \bigg\langle \bigg| \sum_\mu W_{a \mu} [(E - H_1 +
  (i/2) \Gamma_1)^{- 1}]_{\mu N} \sqrt{\lambda} \bigg|^2 \bigg\rangle
\nonumber \\
&& \qquad \qquad = \frac{T_a}{\sum_{a'} T_{a'}} \ .
\ea

Using Eqs.~(\ref{a3}) and (\ref{a5}) and the corresponding results for
$\xi_2$, we can now give the asymptotic forms of the average
transition probability. We distinguish three cases.

(i) The inequality $\sum_{b'} T_{b'} \gg 1$ holds, the variance of
$\xi_2$ is small compared to unity but the variance of $\xi_1$ is
not. Then
\ba
\label{a6}
{\cal A}_{\rm tun} &=& \frac{\tilde{\cal V}}{1 + i \tilde{\cal V} \xi_1
  \tilde{\cal V}} \ , \nonumber \\
{\cal A}_{\rm tra} &=& \tilde{\cal V}_1
\frac{\lambda}{E - E_0 - {\cal V}_1 \xi_1 \tilde{\cal V}_1 + (i/2)
  \Gamma^\downarrow_2} \tilde{\cal V}_2 \ . 
\ea
We define
\ba
\label{a7}
&& P_a = \\
&& \bigg\langle \bigg| \sqrt{2 \pi} \sum_\mu W_{a \mu} [(E - H_1 +
  (i/2) \Gamma_1)^{- 1}]_{\mu N} \sqrt{\lambda} \bigg|^2 | {\cal A} |^2
\bigg\rangle \ . \nonumber
\ea
Here ${\cal A}$ stands for either ${\cal A}_{\rm tun}$ or ${\cal
  A}_{\rm tra}$ defined in Eq.~(\ref{a6}) as the case may be. Then
\ba
\label{a8}
\langle P_{a b} \rangle = \frac{P_a T_b}{\sum_{b'} T_{b'}} \ .
\ea
The ensemble-averaged probability $P_a$ of formation of the transition
channel or transition state is not accessible analytically and can
only be calculated numerically. The normalized decay probability into
channel $b$ is given by $T_b / \sum_{b'} T_{b'}$ and is independent of
the mode of formation of the transition channel or transition state.

(ii) The inequality $\sum_{a'} T_{a'} \gg 1$ holds, the variance of
$\xi_1$ is small compared to unity but the variance of $\xi_2$ is
not. Then
\ba
\label{a9}
{\cal A}_{\rm tun} &=& \frac{\tilde{\cal V}}{1 + i \tilde{\cal V} \xi_2
  \tilde{\cal V}} \ , \nonumber \\
{\cal A}_{\rm tra} &=& \tilde{\cal V}_1
\frac{\lambda}{E - E_0 + (i/2) \Gamma^\downarrow_1 - {\cal V}_2
  \xi_2 \tilde{\cal V}_2} \tilde{\cal V}_2 \ .
\ea
We define
\ba
\label{a10}
&& P_b = \\
&& \bigg\langle | {\cal A} |^2 \bigg| \sqrt{2 \pi} \sum_\nu
\sqrt{\lambda} [(E - H_2 + (i/2) \Gamma_2)^{- 1}]_{(N +1) \nu} W_{\nu b}
\bigg|^2 \bigg\rangle \ . \nonumber
\ea
Here ${\cal A}$ stands for either ${\cal A}_{\rm tun}$ or ${\cal
  A}_{\rm tra}$ defined in Eq.~(\ref{a9}) as the case may be. Then
\ba
\label{a11}
\langle P_{a b} \rangle = \frac{T_a P_b}{\sum_{a'} T_{a'}} \ .
\ea
The ensemble-averaged decay probability $P_b$ in Eq.~(\ref{a10}) is not
available analytically and can only be calculated
numerically. Therefore, Eq.~(\ref{a11}) does not provide a useful
parametrization of the reaction $a \to b$. It does, however, usefully
parametrize the converse reaction $b \to a$ defined below
Eq.~(\ref{s4}). For that reaction, the normalized probability to
populate final channel $a$ is given by $T_a / \sum_{a'} T_{a'}$,
irrespective of the mode of formation of the transition channel or
transition state.

(iii) The inequalities $\sum_{a'} T_{a'} \gg 1$ and $\sum_{b'} T_{b'}
\gg 1$ both hold, the variances of $\xi_1$ and $\xi_2$ are both small
compared to unity. Then
\ba
\label{a12}
{\cal A}_{\rm tun} &=& \frac{\tilde{\cal V}}{1 + \tilde{\cal V}^2} \ ,
\nonumber \\
{\cal A}_{\rm tra} &=& \tilde{\cal V}_1
\frac{\lambda}{E - E_0 + (i/2) \Gamma^\downarrow_1 + (i/2)
  \Gamma^\downarrow_2} \tilde{\cal V}_2
\ea
and
\ba
\label{a13}
\langle P_{a b} \rangle = \frac{T_a}{\sum_{a'} T_{a'}} |{\cal A}|^2
\frac{T_b}{\sum_{b'} T_{b'}} \ .
\ea
Here ${\cal A}$ stands for either ${\cal A}_{\rm tun}$ or ${\cal
  A}_{\rm tra}$ defined in Eq.~(\ref{a12}) as the case may be. The
normalized decay probability into channel $b$ is given by $T_b /
\sum_{b'} T_{b'}$ and is independent of the mode of formation of the
transition channel or transition state. The same statements apply to
the converse reaction defined below Eq.~(\ref{s4}). In contrast to
Eqs.~(\ref{a8}) and (\ref{a11}), Eq.~(\ref{a13}) gives an explicit
expression for the probability of both, the reaction $a \to b$ and
the converse reaction $b \to a$.

\section{Thick Barrier}
\label{thi}

A special case is tunneling through a thick barrier. Then $\tilde{\cal
  V} \ll 1$, suggesting that $\xi_2 \tilde{\cal V} \xi_1 \tilde{\cal
  V}$ in the denominator of ${\cal A}_{\rm tun}$ in the first of
Eqs.~(\ref{s6}) may be neglected. For that to be the case, the square
roots of the variances of $\xi_1$ and $\xi_2$ must be small in
magnitude compared to $1 / \tilde{\cal V}$. Since $1 / \tilde{\cal V}
\gg 1$ that is a much weaker condition than imposed in
Section~\ref{asy} where the square roots had to be small compared to
unity. The condition is fulfilled already when only few channels are
coupled to $H_1$ and to $H_2$. In that case the asymptotic expressions
in Section~\ref{asy} do not apply, and the variances of $\xi_1$ and
$\xi_2$ must be obtained by direct calculation of the threefold
integral in Eq.~(\ref{x2}). If $\xi_2 \tilde{\cal V} \xi_1 \tilde{\cal
  V}$ is indeed negligible, lowest-order perturbation theory in
$\tilde{\cal V}$ is appropriate,
\ba
\label{tb1}
{\cal A}_{\rm tun} = \tilde{\cal V}
\ea
does not fluctuate, $P_{a b}$ in Eq.~(\ref{s8}) is the product of
three statistically uncorrelated factors, and transition-state theory
holds. For $\langle P_{a b} \rangle$ we find
\ba
\label{tb2}
&& \langle P_{a b} \rangle = {\cal A}^2_{\rm tun} \\
&& \times \bigg\langle \bigg| \sqrt{2 \pi} \sum_\mu
W_{a \mu} [(E - H_1 + (i/2) \Gamma_1)^{- 1}]_{\mu N} \sqrt{\lambda} \bigg|^2
\bigg\rangle  \nonumber \\
&& \times \bigg\langle \bigg| \sqrt{2 \pi} \sum_\nu
\sqrt{\lambda} [(E - H_2 + (i/2) \Gamma_2)^{- 1}]_{(N +1) \nu} W_{\nu b}
\bigg|^2 \bigg\rangle \ . \nonumber
\ea
Formation and decay of the tunneling channel are independent
processes. The associated probabilities, given by the ensemble
averages in Eq.~(\ref{tb2}), can be calculated using the threefold
integral~(\ref{x4}) and its analogue for channel $b$. The formal
symmetry of Eq.~(\ref{tb2}) with respect to the interchange $a
\leftrightarrow b$ reflects the symmetry of the scattering matrix in
Eq.~(\ref{s3}).

Eq.~(\ref{tb2}) does not require that either $\sum_{a'} T_{a'} \gg 1$
or $\sum_{b'} T_{b'} \gg 1$ or both. It suffices that the square roots
of both sums are substantially bigger than $\tilde{\cal V}$, the
(small) dimensionless tunneling matrix element through the
barrier. That weaker constraint may be useful in some practical
cases. Indeed, barrier penetration is small for energies far below the
barrier. The number of open channels in either fragment decreases with
decreasing excitation energy, reducing both $\sum_{a'} T_{a'}$ and
$\sum_{b'} T_{b'}$. A likely result is $1 \approx [(\sum_{a'}
  T_{a'})^{1/2}, (\sum_{b'} T_{b'})^{1/2}] \gg \tilde{\cal V}$.

\section{Summary}
\label{sum}

To model a system intrinsically separated by a barrier, we have used
two GOE Hamiltonians (each coupled to open channels) that are coupled
to each other either by a tunneling matrix element, or by an
intermediate transition state. We have studied that model in the
universal limit of large matrix dimension of random-matrix theory.
The transition probability $P_{a b}$ connecting an incoming channel
$a$ that feeds the first GOE Hamiltonian to an outgoing channel $b$
coupled to the second Hamiltonian, is the product of three
statistically correlated factors. These account, respectively, for
formation of, passage through, and decay of the transition channel or
transition state.

A sufficient condition for transition-state theory to hold in its
standard form is that the sum of the transmission coefficients
$T_{b'}$ accounting for decay of the second Hamiltonian into channels
$b'$ obeys $\sum_{b'} T_{b'} \gg 1$. Then the third of the
above-mentioned factors becomes uncorrelated with the first two, the
average transition probability factorizes, the decay of the transition
channel or transition state is independent of its mode of formation,
and the ensemble-averaged transition probability is
\ba
\label{1}
\langle P_{a b} \rangle = \frac{P_a T_b}{\sum_{b'} T_{b'}} \ . 
\ea
The probability $P_a$ of formation of the transition channel or
transition state is a parameter that is analytically available only if
for the entrance channels $a'$ the analogous condition $\sum_{a'}
T_{a'} \gg 1$ holds as well.

Our formulation is symmetric with regard to the interchange $a
\leftrightarrow b$. Therefore, the average probability for the
converse reaction $b \to a$ is given by
\ba
\label{2}
\langle P_{b a} \rangle = \frac{P_b T_a}{\sum_{a'} T_{a'}}
\ea
provided that $\sum_{a'} T_{a'} \gg 1$.

The conditions $\sum_{b'} T_{b'} \gg 1$ for the reaction $a \to b$ and
$\sum_{a'} T_{a'} \gg 1$ for the converse reaction $b \to a$ are
sufficient for the transition-state formulas~(\ref{1}) and (\ref{2}),
respectively, to hold. The weaker conditions $(\sum_{a'} T_{a'})^{1/2}
\gg \tilde{\cal V}$, $(\sum_{b'} T_{b'})^{1/2} \gg \tilde{\cal V}$
suffice for tunneling through a thick barrier with dimensionless
tunneling matrix element $\tilde{\cal V} \ll 1$. Then the average
probability for the reaction $a \to b$ factorizes. The probabilities
for formation and decay of the tunneling channel given in
Eq.~(\ref{tb2}) do not have the simple form $T_c / \sum_{c'} T_{c'}$
but are available in terms of the threefold integral in
Eq.~(\ref{x4}).

{\bf Acknowledgement.} The author is grateful to G. F. Bertsch for
correspondence and to him and to K. Hagino for a clarifying discussion.

\section*{Appendix: Mean Values and Variances}

We follow Ref.~\cite{Ver85}, referring to equations in
Ref.~\cite{Ver85} by a prefix $V$. For the average $\langle \xi_1
\rangle$ we use the steps leading to Eq.~(V.7.7). That gives
Eq.~(\ref{a2}). For the calculation of the variance of $\xi_1$, we
sketch the important steps. According to Eqs.~(V.2.3) and (V.2.4) with
$H \to H_1$ we have $\xi_1 = \lambda D^{- 1}_{N N}$. For the variance
of $\xi_1$ we use Eq.~(V.3.16c) with $E_1 = E_2$ and $\mu(1) = \nu(1)
= N = \mu(2) = \nu(2)$. The ensemble average of the generating
function $Z$ is given in Eq.~(V.7.10) where $\ve = 0$, $c = 1$,
$\tilde{\alpha} = 0$. The graded matrix ${\bf J}$ equals $\{J_{N N}(1)
I(1), J_{N N} I(2) \}$, with $I(1)$ and $I(2)$ defined below
Eq.~(V.7.12). In the limit $N \to \infty$, the graded matrix $\rho$ in
Eq.~(V.7.11) tends to the unit matrix. We use Eq.~(V.5.41) for
$\sigma_G$ and Eq.~(V.5.29) for $T_0$. Following the steps that lead
to Eq.~(V.7.23) we obtain for the variance of $\xi_1$
\ba
\label{x1}
&& \langle |\xi_1|^2 \rangle - 1 = \frac{1}{16} \int {\rm d} \mu(t)
\bigg\{ {\rm trg} (\alpha_1 I(1)) {\rm trg} (\alpha_2 I(2)) \nonumber \\
&& + 8 {\rm trg} ( t_{2 1} (1 + (1/2) \alpha_1)^{1/2} I(1) t_{1 2} (1
+ (1/2) \alpha_2)^{1/2} I(2) \bigg\} \nonumber \\
&& \ \ \times \exp \{ - (1/2) \sum_{a'} {\rm trg} \ln (1 + (1/2) T_{a'}
\alpha_1) \} \ . 
\ea
Here $T_{a'}$ is the transmission coefficient in channel $a'$ defined
in Eq.~(\ref{a3}). Integration over the saddle-point manifold as in
Section~8 of Ref.~\cite{Ver85} gives
\ba
\label{x2}
&& \langle |\xi_1|^2 \rangle - 1 = \frac{1}{2} \int_0^\infty {\rm d}
\lambda_1 \int_0^\infty {\rm d} \lambda_2 \int_0^1 {\rm d} \lambda
\ \mu(\lambda_1, \lambda_2, \lambda) \nonumber \\
&& \ \ \ \times \prod_{a'} \frac{(1 -
T_{a'} \lambda)}{(1 + T_{a'} \lambda_1)^{1/2}(1 + T_{a'} \lambda_1)^{1/2}}
\nonumber \\
&& \ \ \ \times \bigg( (\lambda_1 + \lambda_2 + 2 \lambda)^2
\nonumber \\
&& \ \ \ \ \ + (\lambda_1 (1 + \lambda_1) + \lambda_2 (1 + \lambda_2)
+ 2 \lambda (1 - \lambda) \bigg) \ .
\ea
The integration measure in Eq.~(\ref{x2}) is given by
\ba
\label{x3}
&& \mu(\lambda_1, \lambda_2, \lambda) = \nonumber \\
&& \frac{(1 - \lambda) \lambda |\lambda_1 - \lambda_2|}
{((1 + \lambda_1) \lambda_1 (1 + \lambda_2) \lambda_2)^{1/2} (\lambda +
\lambda_1)^2 (\lambda + \lambda_2)^2} \ .
\ea
For an isolated system (all $T_{a'} = 0$) the integrals in
Eq.~(\ref{x2}) diverge. That is because for $\lambda_1, \lambda_2 \gg
1$ the leading part of the integrand is $| \lambda_1 - \lambda_2 |
\lambda^{- 3}_1 \lambda^{- 3}_2$ times a factor given by either
$\lambda^2_1$ or $\lambda^2_2$ or $\lambda_1 \lambda_2$. Either of the
first two factors yields a diverging integral. To cure the divergence
it suffices that $H_1$ be coupled to a single open channel $a$ as that
gives rise to an additional factor $T^{- 1}_a (\lambda_1 \lambda_2)^{-
  (1/2)}$. However, for the single-channel case the factor $T^{- 1}_a$
and with it, the variance of $\xi_1$ may be arbitrarily large. More
factors of the form $T^{- 1}_{a'} (\lambda_1 \lambda_2)^{- (1/2)}$
arise as the number of open channels increases, showing that the
variance decreases with the number of open channels. That is confirmed
when we apply the asymptotic expansion in inverse powers of $\sum_{a'}
T_{a'}$ of the GOE scattering matrix in Section~4 of Ref.~\cite{Ver86}
to our case. For the right-hand side of Eq.~(\ref{x2}), Eq.~(4.8) of
that paper yields, in leading order, Eq.~(\ref{a4}).

We turn to the amplitude of formation of the tunneling channel or
transition state. The ensemble average of that amplitude vanishes for
$N \to \infty$. That is seen by following the steps that lead to
Eq.~(V.7.7). The calculation of the variance proceeds along the
lines that lead to Eq.~(\ref{x2}). It does not seem necessary to give
the steps in detail. We find
\ba
\label{x4}
&& 2 \pi \bigg\langle \bigg| \sum_\mu W_{a \mu} [(E - H_1 +
  (i/2) \Gamma_1)^{- 1}]_{\mu N} \sqrt{\lambda} \bigg|^2 \bigg\rangle
\nonumber \\
&=& \frac{T_a}{2} \int_0^\infty {\rm d}
\lambda_1 \int_0^\infty {\rm d} \lambda_2 \int_0^1 {\rm d} \lambda
\ \mu(\lambda_1, \lambda_2, \lambda)  \\ && \times \prod_{a'}
\frac{(1 - T_{a'} \lambda)}{(1 + T_{a'} \lambda_1)^{1/2}(1 + T_{a'}
  \lambda_1)^{1/2}} \nonumber \\
&& \times \bigg( \frac{\lambda_1 (1 + \lambda_1)}{(1 + T_a \lambda_1)}
+ \frac{\lambda_2 (1 + \lambda_2)}{(1 + T_a \lambda_2)}
+ \frac{2 \lambda (1 - \lambda)}{(1 - T_a \lambda)} \bigg) \ . \nonumber
\ea
If $\sum_{a'} T_{a'} \gg 1$, we approximate the right-hand side of
Eq.~(\ref{x4}) by the first term of the asymptotic expansion in
inverse powers of $\sum_{a'} T_{a'}$. Eq.~(4.8) of Ref.~\cite{Ver86}
then gives Eq.~(\ref{a5}). Corresponding expressions hold for the last
factor on the right-hand side of Eq.~(\ref{s8}).

\end{document}